\documentclass[twocolumn]{jpsj3} 

\usepackage{color}
\usepackage{graphicx}
\usepackage{subfigure}

\def\nle{\ \raise.3ex\hbox{$<$}\kern-0.8em\lower.7ex\hbox{$\sim$}\ }
\def\nge{\ \raise.3ex\hbox{$>$}\kern-0.8em\lower.7ex\hbox{$\sim$}\ }

\title{Position of Effective Spins Induced by Dilution in Two-Dimensional Spin-Peierls Systems}

\author{Chitoshi \textsc{Yasuda}$^{1}$\thanks{E-mail address:cyasuda@sci.u-ryukyu.ac.jp} and Shouta \textsc{Miyara}$^{2}$}

\inst{$^{1}$Faculty of Science, University of the Ryukyus, Nishihara, Okinawa 903-0213, Japan \\
$^{2}$Graduate School of Engineering and Science, University of the Ryukyus, Nishihara, Okinawa 903-0213, Japan}

\abst{
The site- and bond-dilution effects of the nonmagnetic ground state of a two-dimensional $S=1/2$ antiferromagnetic Heisenberg model, 
coupled with the lattice distortions on a square lattice, are investigated by performing quantum Monte Carlo simulations. 
In the nondiluted system, a phase diagram parameterized by the interchain interaction and the elastic constant is obtained, and the values of the lattice distortions in the dimerized phase are evaluated precisely.
In the diluted system, we compare two ground-state energies assuming two patterns of lattice distortions with magnetic moments (effective spins) induced near the diluted parts and induced at the midpoint between the diluted parts. As a result, we find that it is difficult to induce effective spins near diluted parts for large elastic constants, small interchain interactions, and large concentrations of dilution.
}

\recdate{\today}

\begin{document}
\maketitle

\section{Introduction} 

The spin-Peierls transition observed in quasi-one-dimensional antiferromagnets is an interesting physical property that can show quantum effects conspicuously in quantum spin systems. In particular, since the synthesis of a single crystal of the first inorganic spin-Peierls compound CuGeO$_3$~\cite{hase}, the elucidation of the physical properties has advanced greatly. 
In the spin-Peierls compounds, the positions of the magnetic atoms are distorted alternately at low temperatures, and two adjacent magnetic atoms form a spin singlet. The phase transition occurs when the decrease in the energy of the quantum spin systems, originating from lattice distortions, exceeds the increase in the elastic energy~\cite{pytte, cross, nakano}.
The synthesis of the inorganic spin-Peierls compounds has also promoted the research on impurity effects~\cite{martin,masuda,manabe}. When nonmagnetic impurities such as Zn or Mg are substituted for Cu atoms in CuGeO$_3$ (site dilution), an antiferromagnetic long-range order  (AFLRO) is induced by the infinitesimal concentration of dilution at low temperatures. The mechanism of the impurity-induced AFLRO can be understood in terms of an effective spin induced near an impurity~\cite{sigrist, imada,wessel,yasuda}. By substituting a nonmagnetic atom for a magnetic atom, a spin that formed a singlet pair with a removed spin becomes nearly free. We call this an effective spin. A long-range order is induced by the interactions between the effective spins in a sea of spin-singlet pairs.

Such an impurity-induced AFLRO is also observed in bond-disorder systems such as CuGe$_{1-x}$Si$_x$O$_3$, in which the strengths of the interactions between the Cu atoms change randomly~\cite{regnault, masuda2}. Bond-dilution effects in the two-dimensional antiferromagnetic (AF) Heisenberg model consisting of bond-alternating chains have been studied by performing quantum Monte Carlo (QMC) simulations~\cite{yasuda2, yasuda3, yasuda4}. When a bond is removed in the dimerized state, effective spins are induced around two sites at the edges of the removed bond. There are two effective interactions between the effective spins. The AFLRO does not appear until the magnitudes of the two interactions become comparable~\cite{yasuda4}.

In such diluted systems, the effective spins are induced near the diluted sites or bonds. However, experimental results that contradict the numerical results have been reported for the bond-disorder system of CuGe$_{1-x}$Si$_x$O$_3$.  From Cu nuclear quadrupole resonance (NQR) data, it was concluded that effective spins are not induced near the impurities~\cite{kikuchi}. As mentioned above, the theoretical studies on dilution effects in the two-dimensional AF Heisenberg model consisting of bond-alternating chains concluded that effective spins were induced near diluted parts~\cite{yasuda, yasuda4}; these studies, however, assumed that the pattern of lattice distortion was the same as that in a nondiluted system up to a certain impurity concentration. Thus, we need to consider the lattice degree of freedom to investigate the effects of the positions of the induced effective spins on the AFLRO.

The lattice degree of freedom has been considered in several theoretical works. A QMC study on an odd-size AF Heisenberg chain model coupled with the lattice degree of freedom found that a magnetic moment was induced at the center of the finite chain~\cite{Hansen1999, onishi}. Two-dimensional systems with various interchain interactions were also investigated~\cite{Hansen1999, Augier1999, Dobry1999, Laflorencie2003, Hara2005}. In each case, the interchain interactions were treated within the mean-field approximation and, as a result, the effective spins were induced near the impurities. In order to reproduce the result obtained from NQR data for CuGe$_{1-x}$Si$_x$O$_3$~\cite{kikuchi}, we must improve the treatment of the interchain interactions.

In this work, we investigate a two-dimensional system, considering not only the lattice degree of freedom but also the interchain interactions, and treat the inter- and intrachain interactions on an equal footing. First, we investigate the lattice distortion and the ground-state phase diagrams of the nondiluted system by performing QMC simulations with the continuous-imaginary-time loop algorithm~\cite{MC,beard,todo}. Furthermore, by assuming two patterns of lattice distortions in which effective spins are induced and not induced near the diluted parts, we investigate the site- and bond-dilution effects of the dimerized ground state by performing QMC simulations.

The remainder of this article is organized as follows. In Sect.~2, we introduce the Hamiltonian and explain the dimerization of the ground state in the nondiluted system. In Sect.~3, we discuss the site- and bond-dilution effects of the dimerized ground state. Finally, we devote Sect.~4 to a summary.

\section{Model and Dimerization of the Nondiluted System} 

\begin{figure}[t]
\centering
	\subfigure[columnar type]
	{\includegraphics[width=0.3\textwidth]{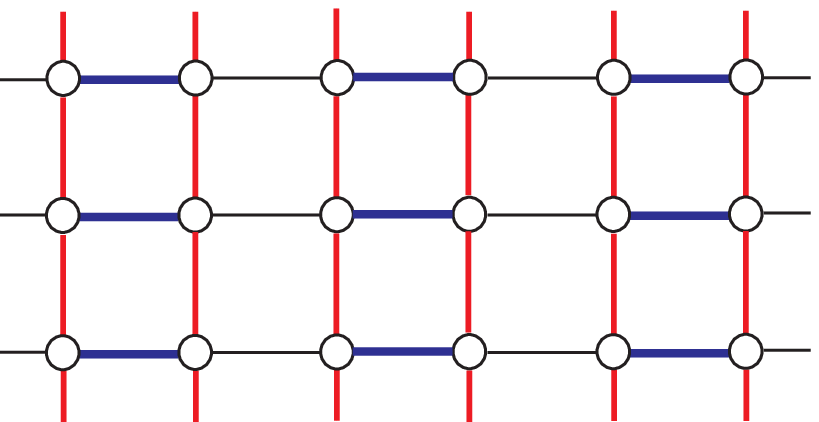}} 	\hspace{1cm}
	\subfigure[staggered type]
	{\includegraphics[width=0.3\textwidth]{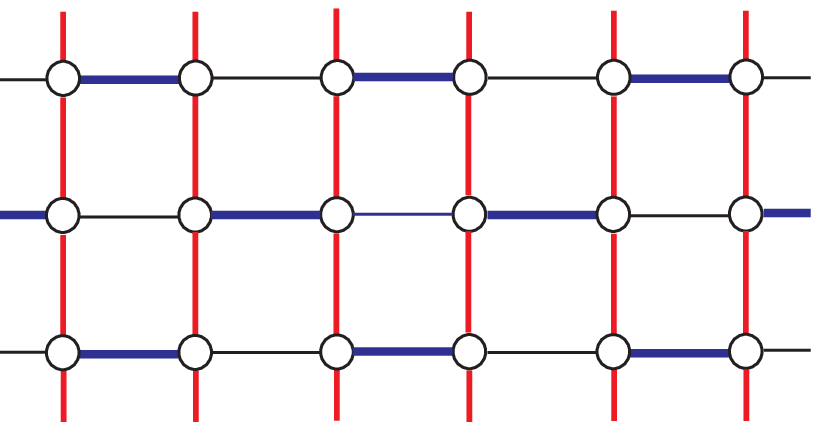}}
  \caption{Illustration of bond alternations in the nondiluted system. Three bond-alternated chains are stacked and the thickness of the bonds depicts the strength of the interactions. The dotted lines express the interchain interactions. The upper figure (a) shows the columnar type and the lower figure (b) shows the staggered type.}  
\label{lattice}
\end{figure}

\begin{figure}[t]
   \centerline{\resizebox{0.45\textwidth}{!}{\includegraphics{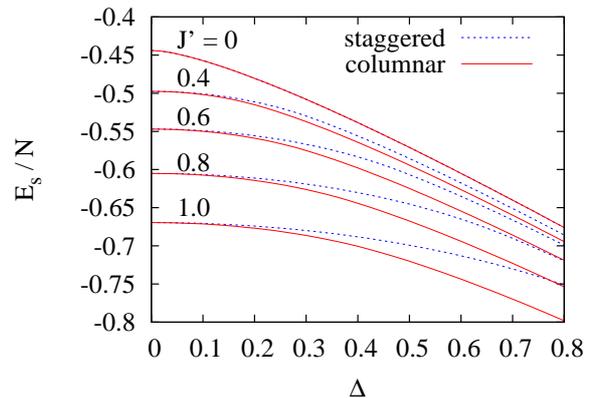}}}
  \caption{Dependences of the ground-state spin energy per site $E_{\rm s}/N$ on the lattice distortion $\Delta$ for the interchain interactions $J' =0$, 0.4, 0.6, 0.8, and 1 in the columnar- and staggered-type lattice distortions.}  
\label{staggered}
\end{figure}

The Hamiltonian of the nondiluted system is described by
\begin{equation}
   {\cal H} = {\cal H}_{\rm sp} + {\cal  H}_{\rm s} + {\cal H}_{\rm p} \ ,
   \label{ham}
\end{equation}
\begin{equation}
 {\cal H}_{\rm sp} = J \sum_{ij}(1+\Delta_{ij}) \mib{S}_{ij} \cdot \mib{S}_{i+1,j}  \ ,
\end{equation}
\begin{equation}
 {\cal H}_{\rm s} = J' \sum_{ij} \mib{S}_{ij} \cdot \mib{S}_{i,j+1}  \ ,
\end{equation}
\begin{equation}
 {\cal H}_{\rm p} = \frac{K}{2}\sum_{ij} \Delta_{ij}^2 \ ,
\end{equation}
where $\mib{S}_{ij}$ is the $S=1/2$ spin operator at site $(i,j)$ on a square lattice with the periodic boundary condition, and $\Delta_{ij}$ describes the lattice distortion between sites $(i,j)$ and $(i+1,j)$. The first term on the right-hand side of Eq.~(\ref{ham}) is the intrachain spin interaction with a spin-lattice coupling within the adiabatic approximation. The strength of $J$ is used as a unit of energy, i.e., $J=1$. The second term is the interchain spin interaction with the exchange integral $J'$ and the third term is the elastic energy for an elastic constant $K$. In the one-dimensional system, the ground state is a dimerized state with bond alternation $\Delta_{ij}=(-1)^i \Delta$, where $\Delta$ is the strength of the lattice distortion. In the two-dimensional system, on the other hand, 
there are two possible candidates for the lattice distortion of the ground state: the columnar-type lattice distortion $\Delta_{ij}=(-1)^i \Delta$, as shown in Fig.~\ref{lattice}(a), and the staggered-type lattice distortion $\Delta_{ij}=(-1)^{(i+j)} \Delta$, as shown in Fig.~\ref{lattice}(b). If $\Delta=1$, the columnar-type and staggered-type lattices become ladder and honeycomb lattices, respectively. Comparing the ground-state spin energies per site $E_{\rm s}/N$ in the range of $0 \le \Delta \le 0.8$, we find that the state with the columnar-type lattice distortion is always stable for $0 < J' \le 1.0$, as shown in Fig.~\ref{staggered}, where $N$ is the number of sites. Thus, we investigate the ground-state phase diagram for the system with the columnar-type lattice distortion. 

\begin{figure}[t]
\begin{center}
 \centerline{\resizebox{0.45\textwidth}{!}{\includegraphics{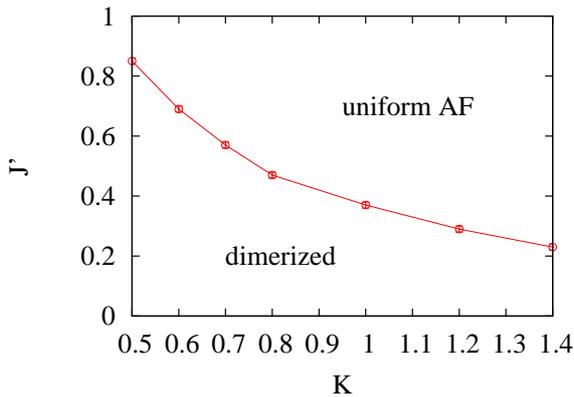}}}
 \caption{Ground-state phase diagram parameterized by the elastic constant $K$ and the interchain coupling $J'$ 
  in the nondiluted system. The solid line passing through the data points is a guide to the eyes.}
\label{pure-phase}
\end{center}
\end{figure}

The ground-state phase diagram estimated from the QMC simulations is shown in Fig.~\ref{pure-phase}. The phase-transition points for each $K$ are estimated from the values of $J'$ at which $\Delta$ becomes zero. 
The values of $\Delta$ are evaluated by the equation
\begin{equation}
    \Delta = \frac{1}{2K} ( \langle \mib{S}_{1,j} \cdot \mib{S}_{2,j} \rangle -  \langle \mib{S}_{2,j} \cdot \mib{S}_{3,j} \rangle )~,
    \label{eq:delta-core}
\end{equation}
which is obtained from the equilibrium condition
\begin{equation}
  \frac{\partial \langle {\cal H} \rangle} {\partial \Delta_{ij}} = 0~,
  \label{eq:eq_condition}
\end{equation}
where the bracket $\langle \cdots \rangle$ in Eqs.~(\ref{eq:delta-core}) and (\ref{eq:eq_condition}) denotes the thermal average.
The constraint $\sum_i \Delta_{ij} = 0$ is satisfied for each chain because $\Delta_{ij}=(-1)^i \Delta$. 
The simulations are carried out at a sufficiently low temperature of $T=0.01$ on sufficiently large square lattices of $32 \times 32$ sites. The physical quantities under consideration in our study do not show any temperature or size dependence and thus, can be identified with those at the ground state within the thermodynamic limit. 
The label `uniform AF' in Fig.~\ref{pure-phase} represents the uniform phase without lattice distortion and with 
an AFLRO, and the label `dimerized' represents the dimerized phase with lattice distortion
and without an AFLRO. Thus, there is no phase in which lattice distortions and an AFLRO coexist in the nondiluted system
for the present accuracy. 
The uniform AF state is stable for large $K$, and the dimerized state is stable for small interchain interactions.
Some of the phase-transition points in Fig.~\ref{pure-phase} have already been calculated by the mean-field approximation and by QMC simulation, and agree with our results~\cite{terai}.

\begin{figure}[t]
  \centerline{\resizebox{0.45\textwidth}{!}{\includegraphics{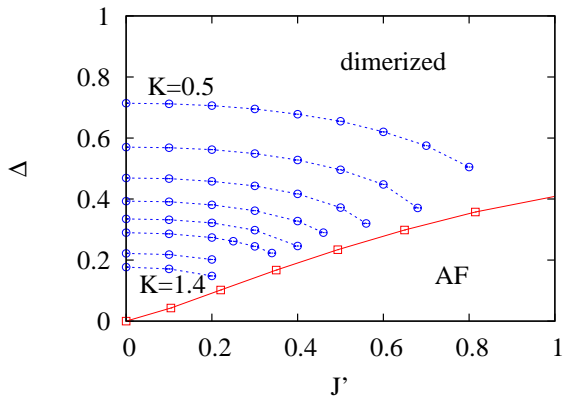}}}
  \caption{Dependences of the lattice distortion $\Delta$ on the interchain interaction $J'$ for various elastic constants $K$ ($K=0.5$, 0.6, 0.7, 0.8, 0.9, 1.0, 1.2, 1.4 from top to bottom) in the nondiluted system. 
  The solid line with squares dividing the dimerized and AF phases shows results obtained from the QMC simulation of the AF Heisenberg model
  with bond alternation in Ref.~28. The solid and broken lines are guides to the eyes.}  
\label{delta-j}
\end{figure}

The $J'$ dependences of $\Delta$ for the dimerized state are shown in Fig.~\ref{delta-j}.
In order to obtain precise values of $\Delta$ to three decimal places, we calculate the correlation functions in Eq.~(\ref{eq:delta-core}) precisely.
That is, the QMC simulations are carried out at a sufficiently low temperature of $T=0.001$ on sufficiently large square lattices of $64 \times 64$ sites.
The dimerized phase survives up to a larger $J'$ for a smaller $K$. The value of $\Delta$ is more sensitive to $K$ than to $J'$.
As $J'$ increases, $\Delta$ rapidly becomes zero at a certain value of $J'$ and a first-order-like phase transition occurs.
The squares in Fig.~\ref{delta-j} are the results of Ref.~28, where the phase-transition line between the AFLRO and dimerized phases was estimated by QMC simulation of the bond-alternated AF Heisenberg model without the elastic energy~\cite{matsumoto, Katoh1994}. This line agrees well with the phase-transition line at which $\Delta$ becomes zero.
This result shows that the lattice distortion becomes uniform with the appearance of the AFLRO.

\begin{figure}[t]
  \centerline{\resizebox{0.45\textwidth}{!}{\includegraphics{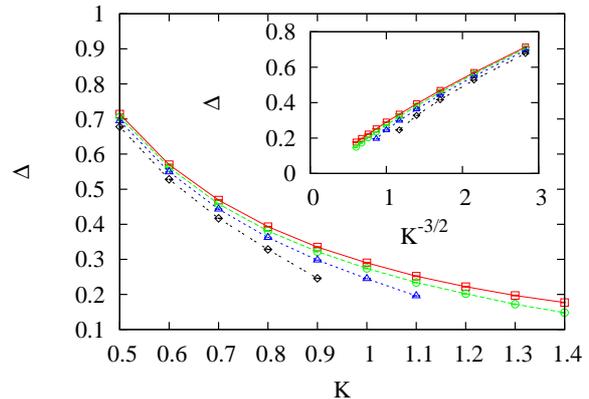}}}
  \caption{Dependences of the lattice distortion $\Delta$ on the elastic constant $K$ for $J'=0$, 0.2, 0.3, and 0.4 expressed by the squares, circles, triangles, and diamonds, respectively. The inset shows the dependences of $\Delta$ on $K^{-3/2}$. All the lines are guides to the eyes.}
\label{delta-k}
\end{figure}
\begin{figure}[t]
  \centerline{\resizebox{0.45\textwidth}{!}{\includegraphics{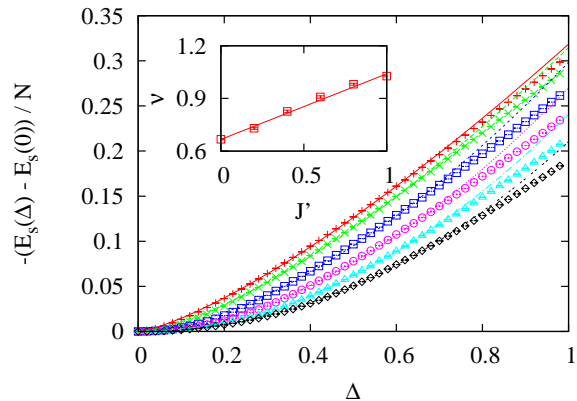}}}
  \caption{Dependences of the spin energy per site on the lattice distortion $\Delta$ for $J'=0$, 0.2, 0.4, 0.6, 0.8, and 1 expressed by the pluses, crosses, squares, circles, triangles, and diamonds, respectively. The lines are the fitting functions $y=ax^{2\nu}$. The inset shows the $J'$-dependence of the exponent $\nu$. The line in the inset is the fitting function $\nu=2/3+0.375J'$.}  
\label{ene-ene0_multi}
\end{figure}

The $K$-dependences of $\Delta$ for various $J'$ are shown in Fig.~\ref{delta-k}. For $J'=0$, the relation $\Delta \sim K^{-3/2}$ is derived from the relation $E_{\rm s}(\Delta)-E_{\rm s}(0) \sim \Delta^{4/3}$ for small $\Delta$, where $E_{\rm s}(\Delta)$ is the ground-state energy of the Hamiltonian ${\cal H}_{\rm sp} + {\cal  H}_{\rm s}$~\cite{cross, nakano}. The same plots with the horizontal axis of $K^{-3/2}$ are shown in the inset of Fig.~\ref{delta-k}. Although we obtain a trend of $\Delta \sim K^{-3/2}$ for $J'=0$, we can observe discrepancies in the trend for finite $J'$.
Since the discrepancy from $\Delta \sim K^{-3/2}$ is caused by that from the relation $E_{\rm s}(\Delta)-E_{\rm s}(0) \sim \Delta^{4/3}$, we show the $\Delta$-dependences of $E_{\rm s}(\Delta)-E_{\rm s}(0)$ in Fig.~\ref{ene-ene0_multi}. The simulations are carried out at a sufficiently low temperature of $T=0.01$ on sufficiently large square lattices of $32 \times 32$ sites.
The condition for which the ground state is the dimerized state is $-(E_{\rm s}(\Delta)-E_{\rm s}(0)) > E_{\rm K}$ for $\Delta \ne 0$, where $E_{\rm K}$ is the ground-state energy of the Hamiltonian ${\cal H}_{\rm p}$. 
For all $J'$, $-(E_{\rm s}(\Delta)-E_{\rm s}(0))$ increases monotonically with $\Delta$, as shown in Fig.~\ref{ene-ene0_multi}. We show the $J'$ dependences of the exponent $\nu$ in the inset of Fig.~\ref{ene-ene0_multi}, assuming that $E_{\rm s}(\Delta)-E_{\rm s}(0) \propto \Delta^{2\nu}$. At $J'=0$, the relation $\nu = 2/3 < 1$ suggests that the ground state of the one-dimensional system becomes the dimerized state since the elastic energy satisfies $E_{\rm K} \propto \Delta^2$. For $J' \ne 0$, on the other hand, the region of the dimerized state is expected to become narrow as $J'$ increases since $\nu$ increases with $J'$ such that $\nu=2/3+0.375J'$. 
The phase transition points shown in Fig.~\ref{pure-phase} are different from those estimated from $\nu=2/3+0.375J' < 1$. This shows that $E_{\rm s}(\Delta)-E_{\rm s}(0)$ is not expressed simply by $\Delta^{2\nu}$ for $J' \ne 0$. The precise estimation of numerically calculated data was not performed in the present work because it deviated from the primary purpose. It is expected that the functional form of energies for the two-dimensional system can be theoretically derived.

\section{Diluted Systems}

In this section, we investigate the dilution effects on the dimerized ground state discussed in the previous section. We fix two diluted sites or bonds on the same chain called the diluted chain. The Hamiltonian of the diluted systems is the same as the Hamiltonian in Eq.~(\ref{ham}) except for the diluted parts. For site and bond dilutions, there is no interaction around diluted sites and on diluted bonds, respectively. 

\subsection{Site dilution} 

\begin{figure*}[t]
\centering
	\subfigure[A-type lattice distortion]
	{\includegraphics[width=0.8\textwidth]{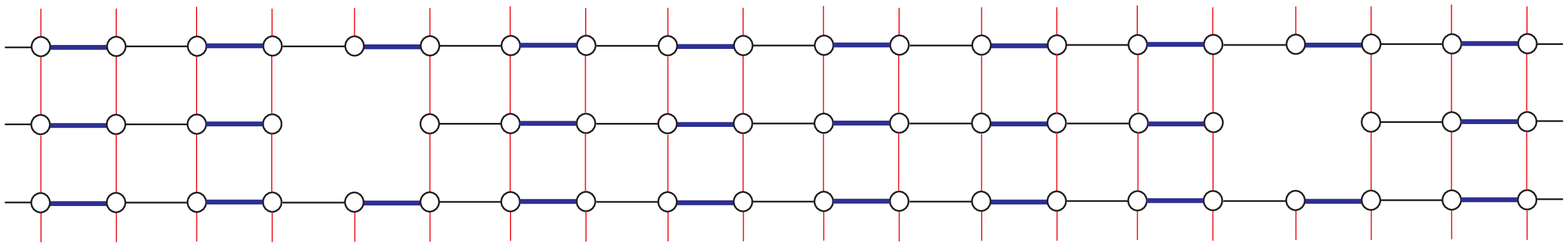}}
	
	\subfigure[B-type lattice distortion]
	{\includegraphics[width=0.8\textwidth]{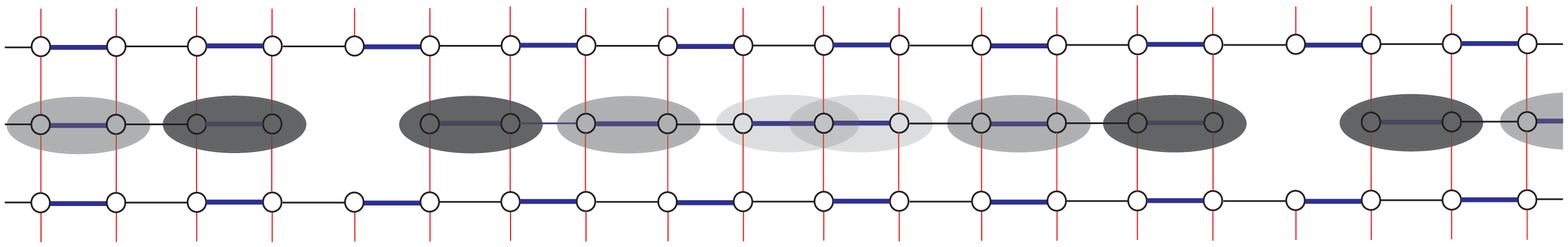}}
  \caption{Illustration of (a) A-type and (b) B-type lattice distortions in the site-diluted system. The thickness of bonds in the chains depicts the strength of the interaction. The bonds surrounded by ellipses are the strong bonds and the intensity of the gray color expresses the strength. While alternation of the nondiluted system is maintained for the A-type, the strengths of the bonds near the diluted sites are the greatest for the B-type.}  
\label{dilution}
\end{figure*}

First, we consider site dilution. In order to investigate the possibility that an effective spin is induced at the midpoint between diluted sites, we consider two types of lattice distortion for the diluted chain, as shown in Fig.~\ref{dilution}. A lattice distortion leads to a change in the strength of the interaction. In Fig.~\ref{dilution}, we express the strength of the interaction caused by a lattice distortion by the thickness of bonds and the intensity of the gray color of the ellipses around the bonds. The A-type lattice distortion is the bond-alternated distortion where the effective spin is induced near the diluted site. The B-type lattice distortion is the distortion experimentally expected in CuGeO$_3$, where the effective spin is induced at the midpoint between the two diluted parts~\cite{kikuchi}. The two types of distortions on the diluted chain $j$ are described by the sinusoidal-type function
\begin{equation}
  \Delta_{ij} = \Delta_{\rm dil} (-1)^{i-1} \cos{\{\frac{\pi m}{N_{\rm b}-1} (i-1) \}} \ ,
  \label{eq:deltaij}
\end{equation}
where $\Delta_{\rm dil}$ is the amplitude of the sinusoidal lattice distortion and $N_{\rm b}$ is the number of bonds between the diluted sites in the diluted chain. The index $m$ denotes the type of distortion: $m=0$ and 1 represent the A-type and B-type, respectively. This sinusoidal-type distortion given by Eq.~(\ref{eq:deltaij}) has been numerically suggested to be realized in an open chain~\cite{Hansen1999, onishi}. Although the hyperbolic tangent function has also been considered as a type of distortion on the diluted chain,~\cite{Saito2002} we use the sinusoidal-type in the present work. Since the A-type is realized in the nondiluted system, the A-type is expected to be realized at the limit of zero concentration of dilution. For the B-type, on the other hand, spin pairs near the diluted sites always form spin-singlet pairs in the diluted chain. Since the B-type is realized in the odd-size open chain~\cite{Hansen1999, onishi}, the B-type is expected to be realized at the limit of zero $J'$.
 In the present work, we assume that the lattice distortions of the nondiluted chain are not affected by dilution and that they are expressed by $\Delta_{ij} = (-1)^i \Delta$, the same as those of the nondiluted system. For the quasi-one-dimensional magnet CuGeO$_3$, where the value of the interchain interaction has been suggested to be $J' \sim 0.1$~\cite{Dobry1999}, this assumption seems to be reasonable.

\begin{figure*}[t]
 \centerline{\resizebox{0.45\textwidth}{!}{\includegraphics{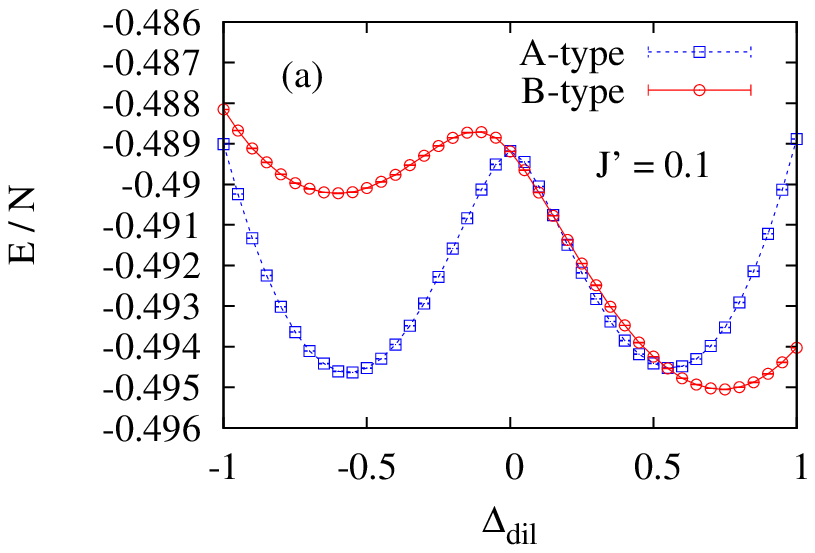}}
 \resizebox{0.45\textwidth}{!}{\includegraphics{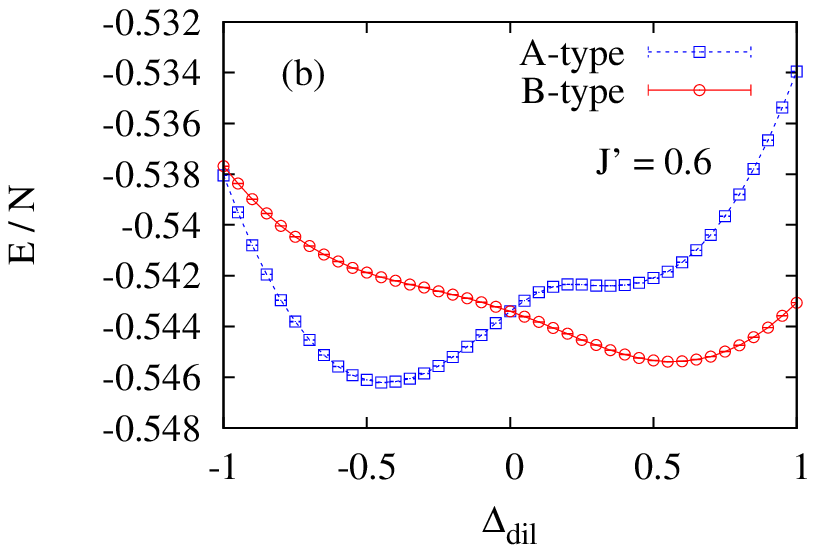}}}
  \caption{Comparison between total energies per site of the systems with the A- and B-types of lattice distortions for (a) $J'=0.1$, (b) $J'=0.6$, and $K=0.6$ in the $L_x \times L_y = 64 \times 8$ system. The B-type and A-type are stable states for $J'=0.1$ and 0.6, respectively. The lines are guides to the eyes.}  
\label{energy}
\end{figure*}

We evaluate the ground-state energies of $L_x \times L_y$-size systems with A- or B-type lattice distortions by performing the QMC simulation. Here, $L_x$ is the length of the spin chain and $L_y$ is the total number of spin chains. The diluted chain has $L_x-2$ sites and there are $L_y-1$ nondiluted chains. We impose the periodic boundary condition. The two diluted spins are arranged such that the diluted chain with the periodic boundary condition is bisected. In Fig.~\ref{energy}, we show examples of the $\Delta_{\rm dil}$ dependence of the total energy, which is the summation of the spin and elastic energies for $K=0.6$ in an $L_x \times L_y = 64 \times 8$ system. We use the results at low temperatures of $T=0.01$ and 0.005 to confirm that there is no temperature dependence. The results in Fig.~\ref{energy} are regarded as those at zero temperature. The lattice distortions of the nondiluted chains are $\Delta=0.568(1)$ and 0.448(1) for $J'=0.1$ and 0.6, respectively. Comparing the results obtained for the A- and B-types, we find that the B-type is stable at $\Delta_{\rm dil}=0.75$ for $J'=0.1$ in Fig.~\ref{energy}(a) and that the A-type is stable at $\Delta_{\rm dil}=-0.45$ for $J'=0.6$ in Fig.~\ref{energy}(b). The minus sign for $\Delta_{\rm dil}$ indicates that the strengths of the bonds on the right-hand side of the diluted sites are weak, as shown in Fig.~\ref{dilution}(a). While the absolute value of $\Delta_{\rm dil}$ for the A-type is close to the value of $\Delta$ in the nondiluted chain, the value of $\Delta_{\rm dil}$ for the B-type is significantly larger than that of $\Delta$. The main cause of the difference might be the use of the sinusoidal-type distortion. If we select the hyperbolic tangent function as the lattice distortion, we predict that the value of $\Delta_{\rm dil}$ will also become close to that of $\Delta$ for the B-type, because the effective spin induced by the hyperbolic-tangent-type distortion is more localized than that induced by the sinusoidal-type distortion. The main terms contributing to the difference in energy between the A- and B-type distortions are the spin energies of bonds in the diluted chain and those of the intrachain bonds connected with the diluted chain, and the elastic energy of bonds in the diluted chain. The magnitude relation of the summation of these terms determines the stable type of lattice distortion. The ground-state phase diagram is determined by such comparisons.

\begin{figure*}[t]
 \centerline{\resizebox{0.45\textwidth}{!}{\includegraphics{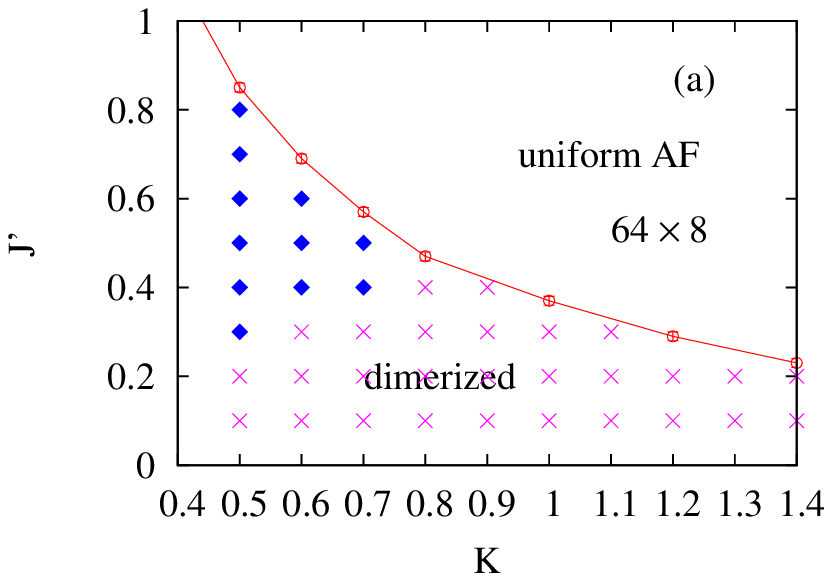}}
 \resizebox{0.45\textwidth}{!}{\includegraphics{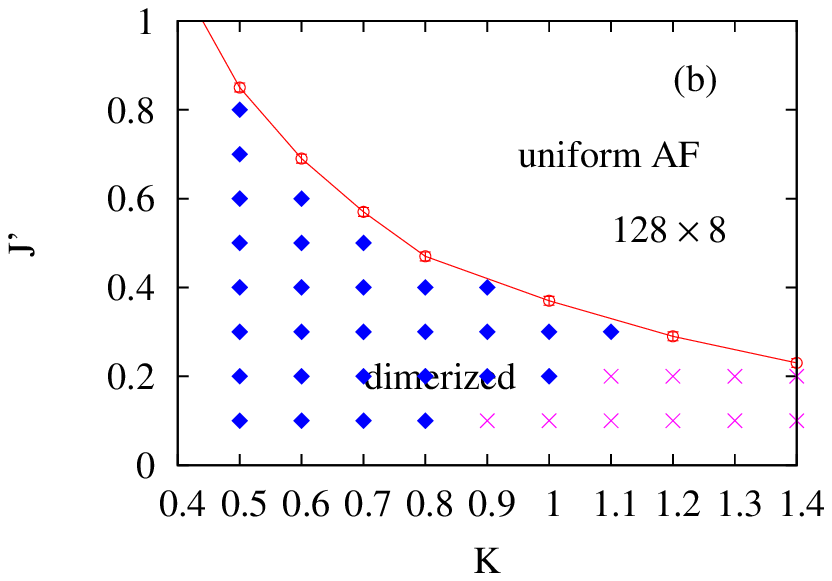}}}
  \caption{Phase diagrams of the site-diluted systems with $L_x \times L_y =$ (a) $64 \times 8$ and (b) $128 \times 8$. Two spins are diluted in the region of the dimerized phase of the nondiluted system. The bold diamonds and the crosses denote the A-type ($m=0$) and B-type ($m=1$)
   lattice distortion states, respectively. The circles with error bars denote the ground-state phase transition line in the nondiluted system shown in Fig.~\ref{pure-phase}. The solid line passing through the circles is a guide to the eyes.}  
\label{2impurity}
\end{figure*}

In Fig.~\ref{2impurity}, we show the ground-state phase diagrams of the site-diluted system in the region where the ground state of the nondiluted system is the dimerized state. Figures \ref{2impurity}(a) and \ref{2impurity}(b) are the phase diagrams for the $L_x \times L_y = 64 \times 8$ and $128 \times 8$ systems, respectively. The difference in $L_x$ corresponds to the concentration of site dilution, because the number of diluted sites is set to a fixed value of 2. The concentration of dilution in Fig.~\ref{2impurity}(a) is larger than that in Fig.~\ref{2impurity}(b). We have confirmed that a further increase in $L_y$ does not influence the phase diagram.
For both Figs.~\ref{2impurity}(a) and \ref{2impurity}(b), the A-type distortion state indicated by bold diamonds becomes the stable ground state for small $K$ and large $J'$, and the B-type indicated by crosses is stable for large $K$ and small $J'$. This shows that effective spins are not induced near the diluted sites for large $K$ and small $J'$. The realization of the B-type distortion states for small $J'$ is consistent with that of the B-type in the limit of $J' \to 0$~\cite{Hansen1999, onishi}. The value of the lattice distortion $\Delta_{\rm dil}$ for the ground state with the A-type distortion is consistent with that of the nondiluted system, except for the sign, while that for the ground state with the B-type distortion does not agree with that of the nondiluted system. For all the parameters we calculated, the values of $\Delta_{\rm dil}$ for the ground state lie in the range of $-1 < \Delta_{\rm dil} < 1$.
On comparing Figs.~\ref{2impurity}(a) and \ref{2impurity}(b), it can be observed that the region of the B-type distortion extends as $L_x$ is decreased, i.e., the concentration of dilution becomes large. Since the ground state of the nondiluted system has A-type lattice distortion, the decrease in the concentration of dilution would lead to an increase in the region of A-type lattice distortion.

\begin{figure*}[t]
 \centerline{\resizebox{0.45\textwidth}{!}{\includegraphics{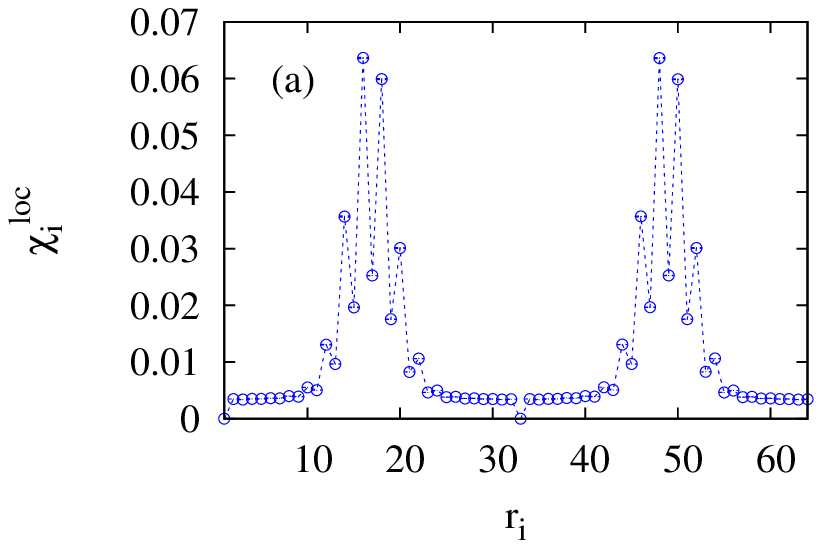}}
 \resizebox{0.45\textwidth}{!}{\includegraphics{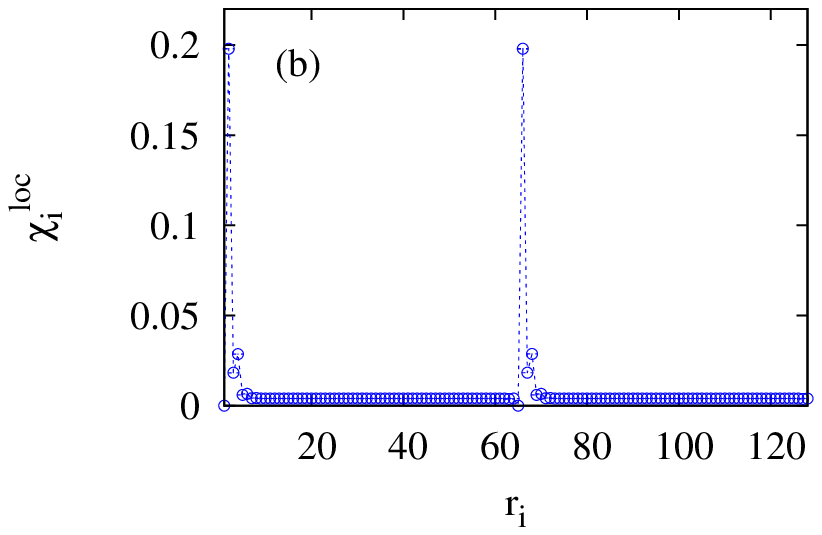}}}
  \caption{Local-field susceptibilities at site $r_i$ for the ground state in the (a) $64 \times 8$ and (b) $128 \times 8$ systems with $K=0.8$ and $J'=0.2$. The lines are guides to the eyes.}  
\label{local-sus}
\end{figure*}

In order to confirm the effective spins actually induced, the local-field susceptibility   
%¡¡
\begin{equation}
  \chi_i^{\rm loc} = \int_0^{\beta}~d\tau \langle S_i^z(0) S_i^z(\tau) \rangle 
\end{equation}
at site $r_i$ on the diluted chain is shown in Fig.~\ref{local-sus}, where $\tau$ is the imaginary time and $\beta$ is the inverse of the temperature. The result in Fig.~\ref{local-sus}(a) is calculated for the B-type ground state with $K=0.8$ and $J'=0.2$ in Fig.~\ref{2impurity}(a). Since the two diluted sites are at $r_i=1$ and 33, the effective spins are induced at the midpoint between the diluted sites. In the case of $L_x=64$ in Fig.~\ref{local-sus}(a), we can expect that a trimer is formed by three spins at the midpoint between the diluted sites, as shown in Fig.~\ref{dilution}(b). Calculating the ground states of the three-spin AF Heisenberg open chain, we find that the magnetic moments on the sites at both ends are larger than that on a site at the center. Such magnetic moments are also seen for the local-field susceptibility in Fig.~\ref{local-sus}(a). Furthermore, the form of the effective spin is asymmetric, i.e.,  the magnitude of the largest $\chi_i^{\rm loc}$ is not the same as that of the second largest $\chi_i^{\rm loc}$. The asymmetry is due to the incommensurability of the pattern of bond alternation in the nondiluted chains next to the diluted chain, as shown in Fig.~\ref{dilution}(b).
The result in Fig.~\ref{local-sus}(b) is calculated for the A-type ground state with $K=0.8$ and $J'=0.2$ in Fig.~\ref{2impurity}(b). The effective spins are induced at the nearest-neighbor sites of the diluted sites $r_i=1$ and 65.

\subsection{Bond dilution} 

\begin{figure*}[t]
\centering
	\subfigure[A-type lattice distortion]
	{\includegraphics[width=0.8\textwidth]{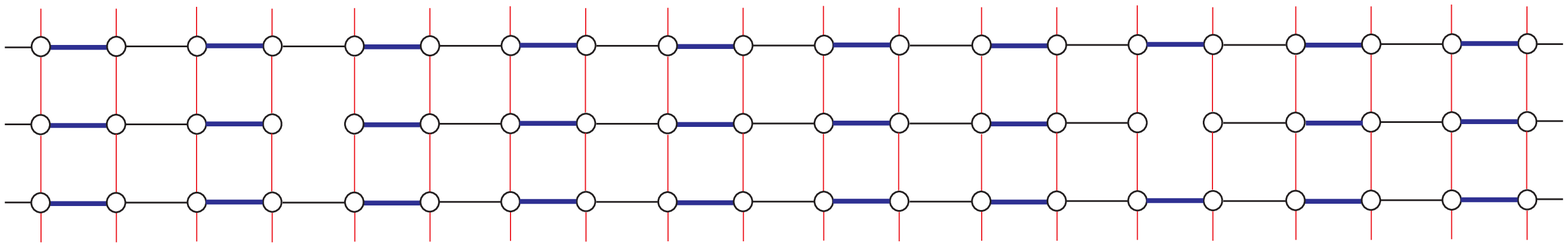}}
	
	\subfigure[B-type lattice distortion]
	{\includegraphics[width=0.8\textwidth]{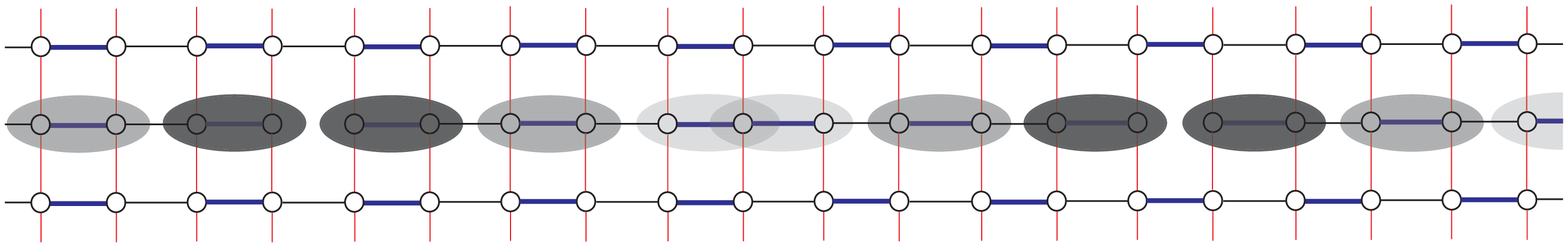}}
  \caption{Illustration of (a) A-type and (b) B-type lattice distortions in the bond-diluted system. The thickness of bonds in the chains depicts the strength of the interaction. The bonds surrounded by ellipses are the strong bonds and the intensity of the gray color expresses the strength. While alternation of the nondiluted system is maintained for the A-type, the strengths of bonds near the diluted bonds are the greatest for the B-type.}  
\label{bond_dilution}
\end{figure*}

\begin{figure*}[t]
 \centerline{\resizebox{0.45\textwidth}{!}{\includegraphics{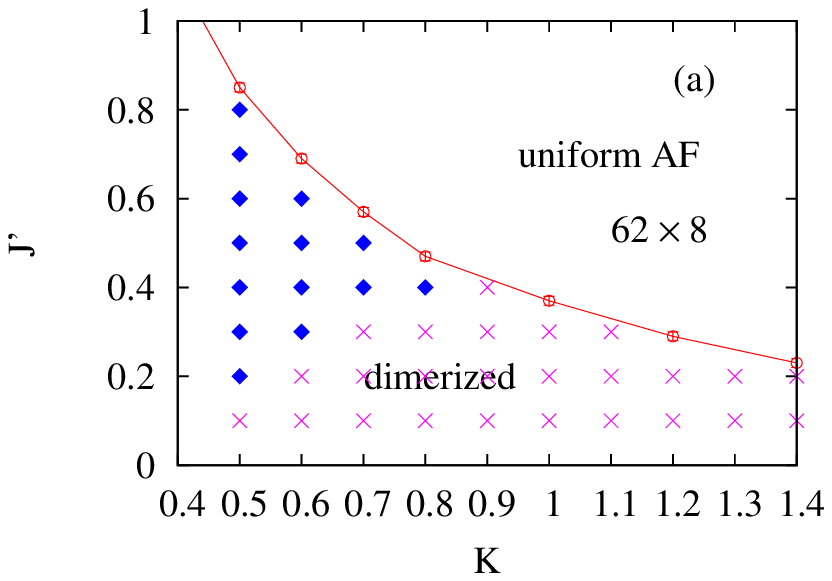}}
 \resizebox{0.45\textwidth}{!}{\includegraphics{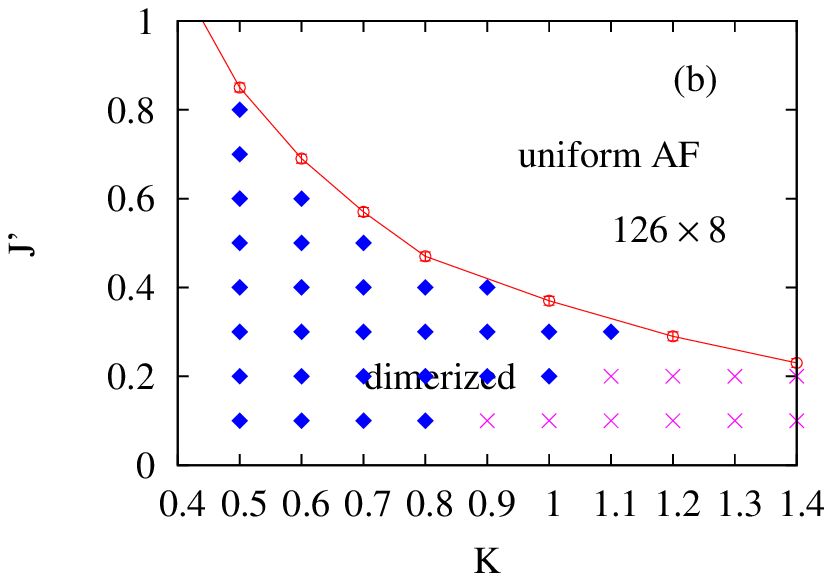}}}
  \caption{Phase diagrams of the bond-diluted systems with $L_x \times L_y =$ (a) $62 \times 8$ and (b) $126 \times 8$. Two bonds are diluted in the region of the dimerized phase in the nondiluted system. The bold diamonds and the crosses denote A-type ($m=0$) and B-type ($m=1$) lattice distortion states, respectively. The circles with error bars denote the ground-state phase transition line in the nondiluted system shown in Fig.~\ref{pure-phase}. The solid line passing through the circles is a guide to the eyes.}  
\label{2impurity-bond}
\end{figure*}

Next, we consider bond dilution. As in the case of site dilution, two types of lattice distortions of the diluted chain are considered, as shown in Fig.~\ref{bond_dilution},  to investigate the possibility that an effective spin is induced at the midpoint between diluted bonds. If the number of spins between diluted bonds is even, the system will always be distorted in the same way as the nondiluted system. Thus, we consider the case where the number of spins is odd. The number of spins in the diluted chain is selected to be the same as that in the site-diluted system.
The A-type distortion is that where the effective spin is induced near the diluted strong bond. The B-type distortion is that where the effective spin is induced at the midpoint between two diluted bonds. In Fig.~\ref{2impurity-bond}, we show the ground-state phase diagrams of the bond-diluted system, where two bonds are diluted. The evaluation method for the stable lattice distortion is the same as that for the site-diluted system. In addition to the phase diagram, the magnitudes of $\Delta_{\rm dil}$ of the stable lattice distortion for each parameter are almost the same as those of the site-diluted system.

The only difference in the phase diagram between the site- and bond-diluted systems is that the A-type region in the bond-diluted system for the $62 \times 8$ lattice, as shown in Fig.~\ref{2impurity-bond}(a), is slightly wider than that in the site-diluted system for the $64 \times 8$ lattice, as shown in Fig.~\ref{2impurity}(a). For example, the results in systems with $(K, J')=(0.5, 0.2)$, (0.6, 0.3), and (0.8, 0.4) are different. As mentioned above, the main terms contributing to the difference in energy between the A- and B-type distortions are the spin energies of bonds in the diluted chain and those of the intrachain bonds connected with the diluted chain, and the elastic energy of bonds in the diluted chain.
Investigating the energies of all nearest-neighbor bonds in the system with different results, we find that the difference in the ground-state energy $E_{\rm s}$ of the Hamiltonian ${\cal H}_{\rm sp} + {\cal H}_s$ between the site- and bond-diluted systems mainly leads to the difference in the phase diagram. The elastic energies do not cause the difference because the number of spins in the diluted chain of the site-diluted system with $L_x=62$ is the same as that of the bond-diluted system with $L_x=64$. In the bond-diluted system with the A-type distortion, the reformation of a spin-singlet pair is known to occur from a previous work on the bond-alternated system~\cite{yasuda4}. As shown in Fig.~\ref{bond_dilution}(a), two effective spins are induced at the ends of the right removed bond, and an AF effective interaction with a strength of $J'^2 (1+\Delta)$ arises between effective spins through the interchain interaction. Therefore, two effective spins reform a spin-singlet pair. As a result, the spin energy of the intrachain bonds connected with the spins at the ends of the removed bond decreases and the spin energy $E_{\rm s}$ of the bond-diluted system with the A-type distortion decreases. As shown in Figs.~\ref{2impurity}(b) and \ref{2impurity-bond}(b), on the other hand, there is no difference in the phase diagram between the site-diluted system for the $128 \times 8$ lattice and the bond-diluted system for the $126 \times 8$ lattice. Since both $J'$ and $\Delta$ are small near the phase boundary in such large systems, the effect of the reformation of the spin-singlet pair on ground-state energies is small. Therefore, it would be difficult to observe a difference in the phase diagram of the large systems.

\section{Summary and Discussion} 

The site- and bond-dilution effects of the nonmagnetic ground state of a two-dimensional $S=1/2$ AF Heisenberg model coupled with the lattice distortions on a square lattice were investigated by performing QMC simulations. In the nondiluted system, a phase diagram parameterized by the interchain interaction and the elastic constant was obtained, and values of the lattice distortions in the dimerized phase were evaluated precisely. The dimerized phase survived up to larger interchain interactions for a smaller elastic constant. As the strength of the interchain interaction increased, the magnitude of the lattice distortion rapidly became zero at a certain value of the interchain interaction and the phase transition occurred.
In the diluted system, by assuming two patterns of lattice distortions with effective spins induced near the diluted parts and induced at the midpoint between the diluted parts, we compared two ground-state energies. We found that it was difficult to induce effective spins near diluted parts for large elastic constants, small interchain interactions, and large concentrations of dilution. 

From our work, we conclude that the position of the induced effective spin depends on the values of the interchain interaction, elastic constant, and concentration of dilution. In the realistic compound CuGeO$_3$, estimating the elastic constant is difficult since a simple adiabatic treatment for the lattice degree of freedom cannot be applied to CuGeO$_3$~\cite{Uhrig1998}. On the other hand, the value of the interchain interaction is estimated to be $J' \sim 0.1$~\cite{Dobry1999}. The concentration of impurities for CuGe$_{1-x}$Si$_x$O$_3$, in which the B-type situation was observed on the basis of NQR data, was approximately 1\%~\cite{kikuchi}. We expect that the A-type situation will be experimentally observed in CuGe$_{1-x}$Si$_x$O$_3$ and Cu$_{1-x}$Mg$_x$GeO$_3$ with low impurity concentrations and in materials having stronger interchain interactions than those in CuGeO$_3$. 

The substitution of Si at the Ge site of CuGeO$_3$ results in a situation close to bond randomness because the strengths of the interactions between the Cu atoms change randomly. For the bond-randomness system, therefore, it will be interesting to perform the same analysis as that for the diluted systems. Even for the one-dimensional system, the A-type distortion is realized for weak randomness.~\cite{Augier2000} On the other hand, the B-type distortion is realized for strong randomness including bond dilution. Investigation of the situations of the two-dimensional system for which the B-type is realized will be an interesting subject of research.

In our work, we fixed two diluted sites or bonds. If many sites or bonds are removed, diluted chains with various lengths will be mixed. Furthermore, we assumed that the lattice distortion of the nondiluted chain was not affected by dilution. In the case where the system has large interchain interactions, we should consider the effects of the lattice distortion of the nondiluted chain on dilution. In future works, we plan to examine these effects.

\vspace*{0.5cm}

\begin{acknowledgments}
The authors acknowledge S. Todo for the stimulating discussions and thank the Supercomputer Center, the Institute for Solid State Physics, the University of Tokyo, for the use of the facilities. This work was supported by JSPS KAKENHI Grant Number JP16K05479. 
\end{acknowledgments}

\end{document}